\documentclass[preprint, 12pts]{elsarticle}
\usepackage{lineno,hyperref}
\usepackage{amsmath}
\usepackage{xspace}
\usepackage{siunitx}
\usepackage{graphicx}
\usepackage{multirow}
\usepackage{booktabs}
\usepackage{makecell}
\usepackage{subcaption}
\usepackage{color}
\usepackage{placeins}
\journal{Nuclear Instruments and Methods A}
\newcommand{\microns}{\si{\micro\meter}\xspace}
\begin{document}
\begin{frontmatter}
\title{Spatial resolution studies with the BabyIAXO Micromegas prototype}

%in prepration
%\author[a,c]{A.~Quintana}
%\author[a]{E.~Ferrer-Ribas}
%\author[a]{A.~Giganon}
%\author[a]{C.~Goblin}
%\author[a]{C.~Loiseau}
%\author[a]{T.~Papaevangelou}

%\author[b]{N.~Goyal}
%\author[b]{F.J.~Iguaz}

%\author[c]{J. Castel}
%\author[c]{S. Cebrián}
%\author[c]{T. Dafni}
%\author[c]{D. Díez-Ibáñez}
%\author[c]{J. Galán}
%\author[c]{J.A. García}
%\author[c]{A. Ezquerro}
%\author[c]{I.G Irastorza}
%\author[c]{G. Luzón}
%\author[c]{C. Margalejo}
%\author[c]{H. Mirallas}
%\author[c]{L. Obis}
%\author[c]{O. Pérez}
%\author[c]{J. Porrón}
%\author[c]{and M. J. Puyuelo}

% new list by alphabetical order

\author[a,b]{A.~Quintana}
\author[b]{J.~Castel}
\author[b]{S.~Cebrián}
\author[b]{T.~Dafni}
\author[b]{D.~Díez-Ibáñez}
\author[a]{E.~Ferrer-Ribas}
\author[b]{A.~Ezquerro}
\author[b]{J.~Galán}
\author[b]{J.A.~García}
\author[a]{A.~Giganon}
\author[a]{C.~Goblin}
\author[c]{N.~Goyal}
\author[c]{F.J.~Iguaz}
\author[b]{I.G~Irastorza}
\author[b]{C.~Loiseau}
\author[b]{G.~Luzón}
\author[b]{C.~Margalejo}
\author[b]{H.~Mirallas}
\author[b]{L.~Obis}
\author[a]{T.~Papaevangelou}
\author[b]{O.~Pérez}
\author[b]{J.~Porrón}
\author[b]{and M.J.~Puyuelo}

\affiliation[a]{organization={IRFU, CEA, Université Paris-Saclay},
            city={Gif-sur-Yvette},
            postcode={91191},
            country={France}}
\affiliation[b]{organization={Centro de Astropartículas y Física de Altas Energías, Universidad de Zaragoza},
            postcode={50009 Zaragoza},
            country={Spain}}
\affiliation[c]{organization={SOLEIL Synchrotron},
            city={L’Orme des Merisiers, Départementale 128},
            postcode={91190 Saint-Aubin},
            country={France}}

\hyphenation{Baby-IAXO}
\begin{abstract}
The spatial resolution of the Micromegas prototype developed for the BabyIAXO experiment was evaluated using a low-energy X-ray beam at the SOLEIL synchrotron facility. BabyIAXO, currently under construction, aims to search for hypothetical solar axions. A key component of the experiment is a low-background X-ray detector with high efficiency in the 1–10\,keV energy range and stringent background rejection capabilities. Achieving a spatial resolution on the order of, or better than, \SI{1}{mm} is critical for accurately reconstructing signal shapes and positions, and for effectively discriminating between signal and background events. Therefore, a precise characterization of the detector's spatial resolution is essential to validate its suitability for the experiment.

This study involved scanning the IAXO-D1 Micromegas detector under various beam energies, positions, and drift field configurations to evaluate their influence on spatial resolution. A resolution of approximately \SI{100}{\micro\meter} at 6\,keV was achieved, confirming the strong potential of this technology for application in the final BabyIAXO setup.

\end{abstract}


\end{frontmatter}
%\linenumbers
%\maketitle
\newpage
%\tableofcontents
%\newpage
%%%%%%%%%%%%%%%%%%%%%%%%%%%%%%%%%%%%%%%%%%%%%%%%%%%%%%
\section{Introduction}
\label{section:Introduction}
%%%%%%%%%%%%%%%%%%%%%%%%%%%%%%%%%%%%%%%%%%%%%%%%%%%%%%

%CAST results\cite{CAST:2004gzq, CAST:2008ixs, CAST:2011rjr, CAST:2017uph, CAST:2024eil}
%IAXO ~\cite{IAXO:2019mpb, Armengaud:2014gea}
%BabyIAXO\cite{IAXO:2020wwp, IAXO:2024wss}
BabyIAXO~\cite{IAXO:2020wwp, IAXO:2024wss} is a fourth-generation helioscope designed to search for hypothetical solar axions, conceived as a first step towards the International AXion Observatory (IAXO). Axions are hypothetical particles predicted by the Peccei–Quinn mechanism, originally proposed to solve the long-standing strong CP problem in the Standard Model (SM) of particle physics. Beyond this, axions are compelling candidates for cold dark matter (DM) and could potentially explain several astrophysical anomalies~\cite{IAXO:2019mpb, Carenza:2024ehj}.

A helioscope typically consists of three main components: a powerful superconducting magnet to induce axion-photon conversion, X-ray optics to focus the resulting photons, and low-background X-ray detectors to image the signal. The detector must combine high efficiency in the 1–10\,keV energy range with excellent background rejection capabilities down to a level of 10$^{-7}
\,\textnormal{counts keV}^{-1} \, \textnormal{cm}^{-2} \, \textnormal{s}^{-1}$.

This paper presents a detailed study of the spatial resolution of the Micromegas~\cite{Giomataris:1995fq} detector developed for the BabyIAXO experiment. Microbulk Micromegas technology~\cite{Andriamonje:2010zz} is especially well suited for such low-background applications, as demonstrated by its performance in the CAST experiment~\cite{CAST:2004gzq, CAST:2008ixs, CAST:2011rjr, CAST:2017uph, CAST:2024eil} and in recent developments towards fulfilling BabyIAXO requirements~\cite{Altenmuller:2024uza}. Fabricated from radiopure materials such as copper and Kapton, these detectors combine low intrinsic background with high energy resolution, achieved through a precisely defined amplification gap determined by the Kapton thickness. In addition, the implementation of a 2D readout anode enhances spatial resolution, further contributing to effective background rejection.

In the BabyIAXO detector, the expected signal is confined to a small region of the detection plane, typically a few millimeters in size~\cite{IAXO:2020wwp}. This spatial confinement allows for powerful background rejection via fiducial cuts but also imposes constraints on the spatial resolution. To accurately reconstruct the signal’s shape and position and to effectively distinguish it from background events, a spatial resolution on the order of, or better than, \SI{1}{mm} is essential. This makes the precise characterization of the detector’s spatial resolution a crucial step in validating its performance for the experiment.
Moreover, in the event of a positive detection, high spatial resolution could contribute to solar studies, particularly by enabling the determination of the solar temperature and Debye screening scale in different layers of the Sun’s interior~\cite{Hoof:2023jol}.

In this work, we present for the first time a dedicated study of the spatial resolution achieved with the 2D readout anode, within the framework of the BabyIAXO detector design.
%%%%%%%%%%%%%%%%%%%%%%%%%%%%%%%%%%%%%%%%%%%%%%%%%%%%%%
\section{The IAXO-D1 detector}
\label{section:IAXO-D1}
%%%%%%%%%%%%%%%%%%%%%%%%%%%%%%%%%%%%%%%%%%%%%%%%%%%%%%
To achieve the target sensitivities of IAXO and BabyIAXO, it is essential to optimize detector efficiency in the energy range below \SI{10}{keV}, while maintaining ultra-low background levels. This goal requires a combination of low-background techniques, including extensive shielding, radiopurity screening of detector components, and advanced event discrimination strategies based on detailed topological information of background events.

The baseline detection technology for BabyIAXO consists of small Time Projection Chambers (TPCs) equipped with two-dimensional Micromegas readouts fabricated using the microbulk technique~\cite{Andriamonje:2010zz}. These detectors have undergone significant low-background development in recent years, particularly within the CAST experiment \cite{Altenmuller:2024uza} by identifying background sources, refining analysis methods and improving the shielding. %Progressive identification and mitigation of background sources responsible for energy deposits in the Region of Interest (RoI), coupled with refined analysis methods and improved shielding, have led to substantial performance gains.
These developments are comprehensively documented in the BabyIAXO Conceptual Design Report~\cite{IAXO:2020wwp}.

\subsection{Detector Description}
\begin{figure}[t]
\begin{center}
\includegraphics[width=0.98\textwidth]{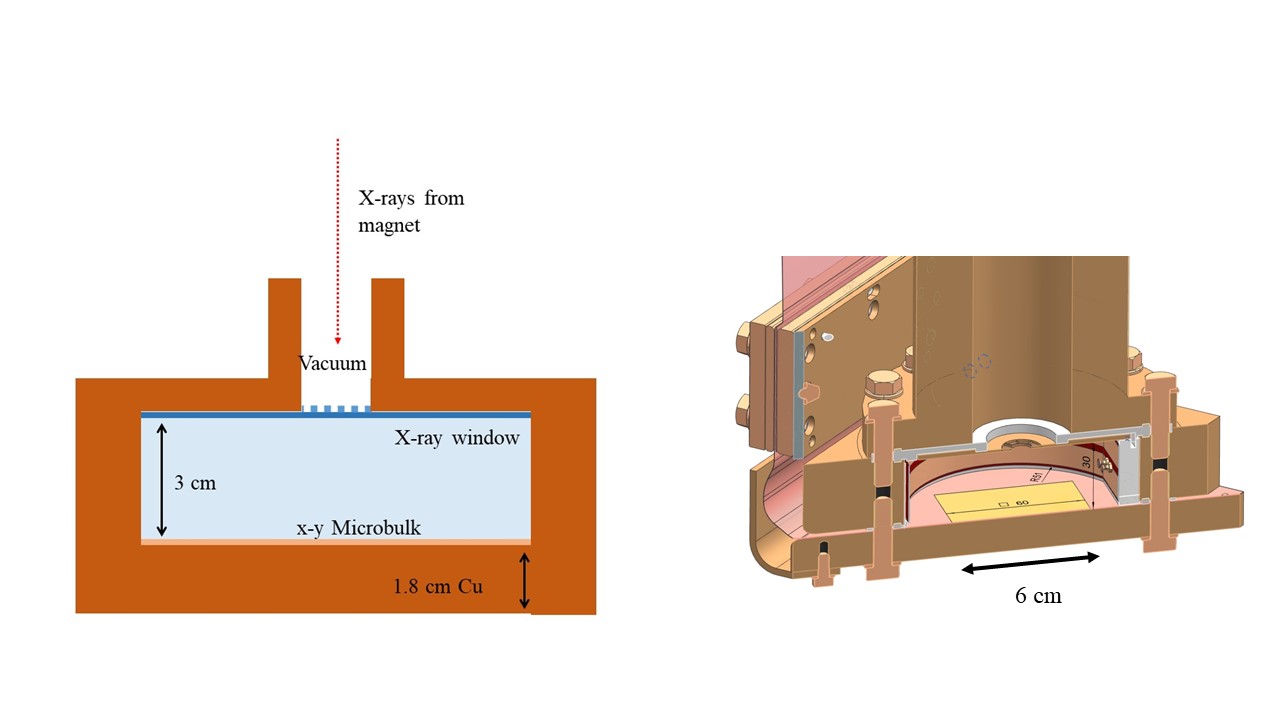}
\caption{\label{fig:MMdesign} Left: Sketch of the design of the Micromegas detector. Not to scale. Right: Cross-section of the  IAXO-D1 prototype.}
\end{center}
\end{figure}

The BabyIAXO Micromegas detector design is based on the last generation of CAST Micromegas detectors \cite{CAST:2024eil, Aznar:2015iia}. The simplified design of the detector is shown in figure~\ref{fig:MMdesign} (left) consisting of a small copper gas chamber with a 2D Micromegas readout plane coupled to the BabyIAXO beam line via an X-ray transparent window.

The detector, named IAXO-D1, is a small TPC with 3\,cm conversion volume filled with argon in addition to a small quantity of quencher (\SI{5}{\%} isobutane) at atmospheric pressure. An alternative gas mixture of \SI{500}{\milli\bar} of Xe (\SI{50}{\%}Xe - \SI{48}{\%}Ne - \SI{2}{\%}Isobutane) is also being considered.
The X-rays coming from the magnet enter the conversion volume via a gas-tight window made of \SI{4}{\micro\m} aluminized mylar foil. This foil is also the cathode of the TPC, and it is supported by a metallic strong-back. The thin windows are designed to withstand the pressure difference between the gas-filled detector and the vacuum line while efficiently transmitting low-energy X-rays (1–10\,keV). 

%ultra thin windows
%The role of the thin windows is to withstand the pressure difference between the detector chamber (filled with gas) and the vacuum line while efficiently transmitting the low-energy X-rays (1-10 keV) to the detector. The thin window foil is supported by a metallic strongback with a spider-web design which leaves a central aperture that conveniently leaves room for the focused spot.

\begin{figure}[t]
\begin{center}
\includegraphics[width=0.8\textwidth]{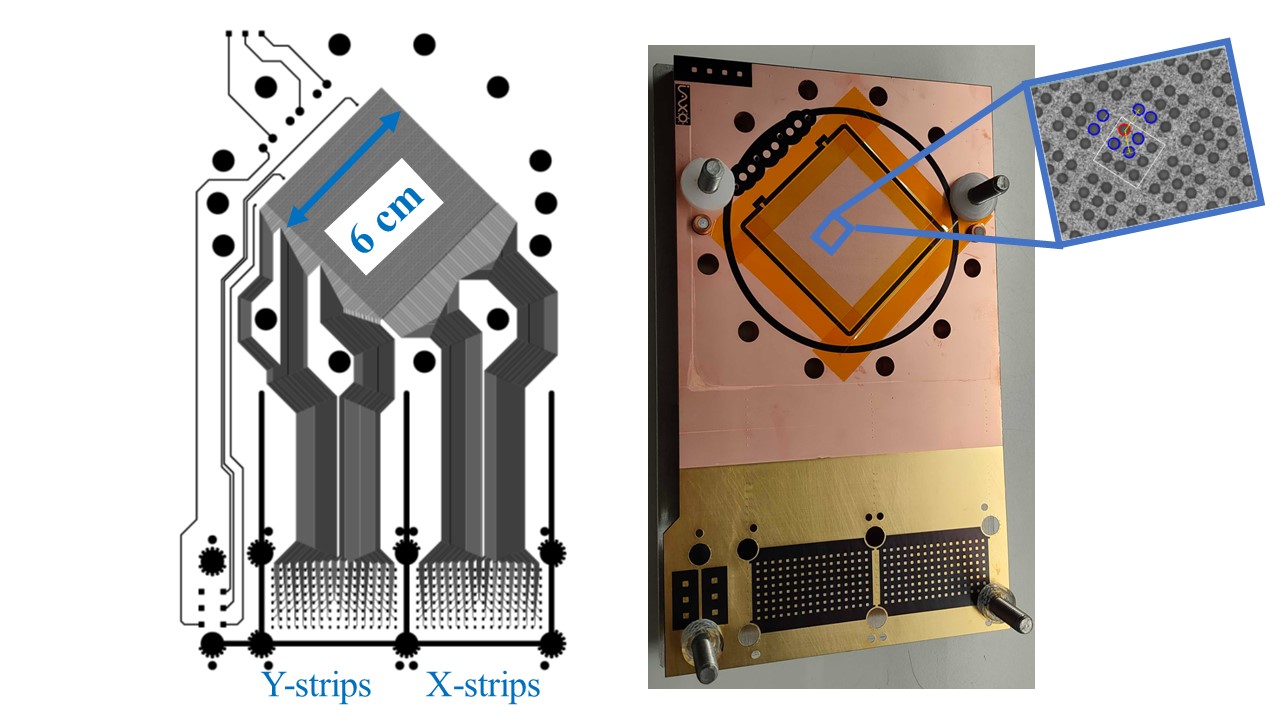}
\caption{\label{fig:microbulk} Left: Readout plane routing of the $6\times6$\,cm$^2$ active area showing the X and Y strips. Right: Picture of the readout plane on its support. The zoom shows a microscope view of the microbulk mesh.}
\end{center}
\end{figure}

The detector chamber is made of \SI{18}{mm} thick radiopure copper (Cu-ETP) walls. A view of the design is shown in figure~\ref{fig:MMdesign} right. All the gaskets are made of radiopure PTFE. A Kapton field shaper has also been installed, to increase the uniformity of the drift field and reduce border effects. The field shaper is externally covered by a \SI{1.5}{mm} thick PTFE coating in order to block the copper fluorescence from the body of the detector. The microbulk-type \cite{Andriamonje:2010zz} readout plane consists of a X-Y strip pattern of 120 strips per axis at a pitch of \SI{500}{\micro\m} covering a surface of $6\times6$\,cm$^2$. In figure~\ref{fig:microbulk} the routing of the readout plane is shown as well as a picture of the Microbulk detector on its support frame. In the zoom, a microscope view of the microbulk mesh is shown where the pattern of the \SI{40}{\micro\m} diameter holes is visible.

\subsection{Front End electronics and acquisition system}
\begin{figure}[t]
\begin{center}
\includegraphics[width=0.6\textwidth]{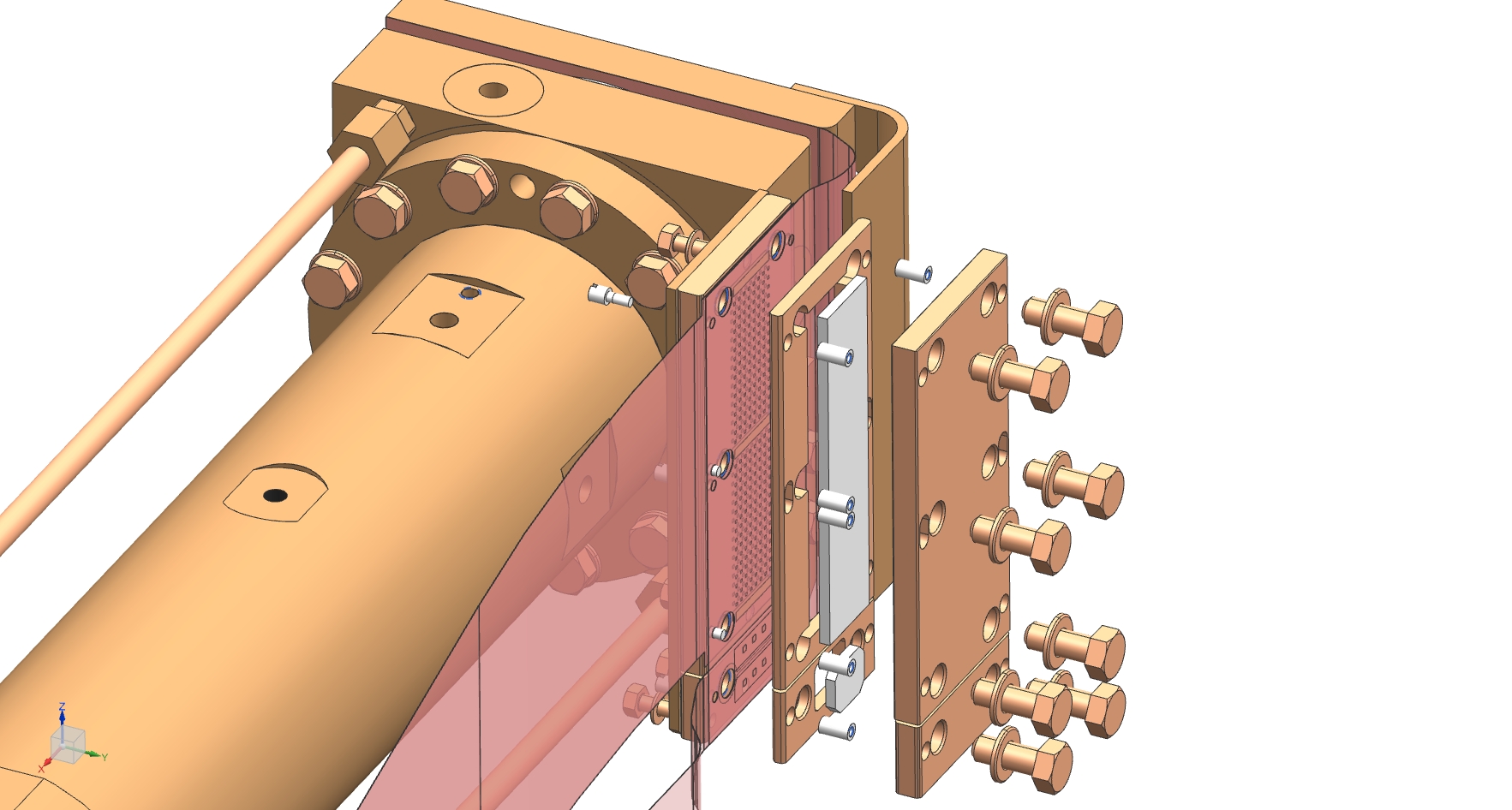}
\caption{\label{fig:face-to-face} Face-to-face connector mechanism for the connection of the readout strips to the front-end electronics.}
\end{center}
\end{figure}

The detector is connected to the front-end electronics via a custom-designed, solder-less interface known as the ``face-to-face connector" \cite{MirallasSanchez:2024bcp}. This system avoids traditional soldering by mechanically compressing a flat Kapton cable with copper pads matching the readout plane pads—between two screwed copper pieces. An illustration of this connector is shown in figure~\ref{fig:face-to-face}. For the present test, we employed the AGET electronics~\cite{Baron:2017kld}, interfaced through a FEMINOS card~\cite{Baron:2017kld}, which acts as a bridge between the analog front-end and the digital data acquisition (DAQ) system by digitizing and aggregating signals from the AGET ASIC. The face-to-face connector is integrated into a 45\,cm-long flat Kapton cable, enabling a direct and reliable connection between the detector and the AGET front-end card. The capacitance of the detector strips, including the contribution from the flat Kapton cable, was measured to be approximately 20\,pF.

\subsection{Implementation for BabyIAXO: lead shielding and cosmic veto}

In its final implementation within the BabyIAXO experiment, the detector will be enclosed by a passive shield consisting of a \si{20} {cm} lead wall designed to suppress external radiation. Surrounding this shielding, an active muon veto system will provide nearly 4$\pi$ coverage with a targeted efficiency of 99\%. This system will use \si{5} {cm}-thick plastic scintillators, each approximately \si{20} {cm} wide and \si{1} {m} long, arranged in a customized geometry. 

Recent results from prototype tests and simulations~\cite{Altenmuller:2024uza} suggest that cosmic-ray–induced neutrons—generated by nuclear interactions of cosmic rays with the atmosphere—may represent a significant background source. These neutrons can interact directly with the detector gas, producing nuclear recoils that mimic the signature of X-rays~\citep{ElisaThesis}.

To further mitigate this background, a neutron tagger is under development. One proposed design consists of three layers of plastic scintillator panels interleaved with cadmium sheets. Monte Carlo simulations indicate that primary neutrons can interact within this setup, producing secondary neutrons that become thermalised in the plastic scintillators. If these thermal neutrons are subsequently captured by the cadmium sheets, characteristic gamma rays are emitted and can be detected by the surrounding scintillators, thereby tagging the neutron interactions.

For the purposes of this test, neither the lead shielding nor the cosmic veto system were included, as they are not pertinent for spatial resolution studies.

%%%%%%%%%%%%%%%%%%%%%%%%%%%%%%%%%%%%%%%%%%%%%%%%%%%%%%
\section{Experimental setup}
\label{section:Setup}
%%%%%%%%%%%%%%%%%%%%%%%%%%%%%%%%%%%%%%%%%%%%%%%%%%%%%%%
The detector was installed in the Metrology beamline \cite{Metrology1, Metrology2} at the SOLEIL synchrotron facility (Saint-Aubin, France), which produces an X-ray beam from 6\,keV to 28\,keV with small divergence and high flux. The beam is shaped using focusing mirrors and collimating slits, while a monochromator is used to tune the energy.

\begin{figure}[h]
    \centering
    \includegraphics[width=1.05\linewidth]{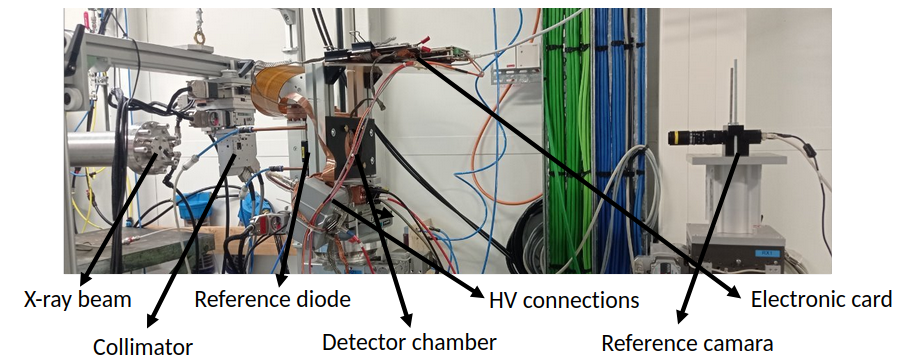}
    \caption{Display of the setup at SOLEIL.}
    \label{fig:setup}
\end{figure}
%\begin{figure}[h]
%    \centering
%    \includegraphics[width=0.8\linewidth]{soleil_setup_front.png}
%    \caption{}
%    \label{fig:setup_front}
%\end{figure}
The setup, shown in figure~\ref{fig:setup}, consists of collimator slits aligned with the beam exit and two movable platforms. On the platform closest to the beam exit, the IAXO-D1 detector was mounted and a diode to use for reference was placed alongside it. On the second platform, a Basler camera \cite{baslerCam} was installed to monitor shape and size of the X-ray beam.

Throughout the test, the detector operated with a gas mixture of argon and 5\% isobutane in an open-loop configuration and maintained a constant flow rate of \SI{5}{L/h}. The mesh voltage was set to \SI{360}{V}, a value chosen to ensure optimal Micromegas performance in terms of both gain ($\sim$\,$\text{10}^\text{4}$) and energy resolution ($\sim$\,18\,\% at 6\,keV). The field shaper was intentionally left disconnected, as its connection was found to introduce significant noise into the data acquisition system.

\begin{figure}[b]
  \centering
  \begin{minipage}[b]{0.47\textwidth}
    \includegraphics[width=5.8cm]{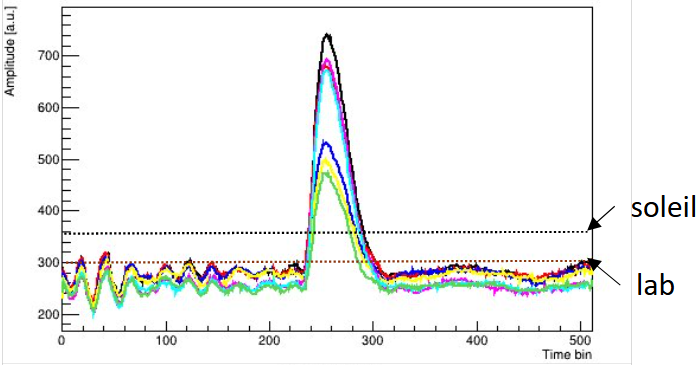}
  \end{minipage}
  \hfill
  \begin{minipage}[b]{0.47\textwidth}
    \includegraphics[width=5.8cm]{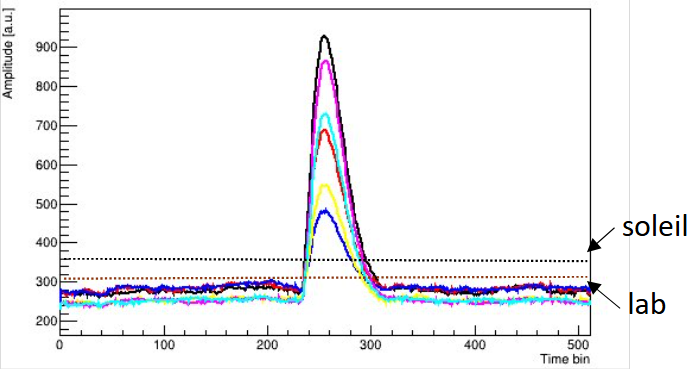}
  \end{minipage}
  \caption{Example of two events with the corresponding raw signals recorded by the detector. The threshold at SOLEIL and laboratory conditions are represented.}
  \label{fig:rarSignals}
\end{figure}

Upon installation of the detector on the beamline, elevated noise levels were recorded in comparison to those observed under controlled laboratory conditions (see figure \ref{fig:rarSignals}). This required the use of an energy threshold of around 120\,ADCs over the baseline, notably higher than the 40\,ADCs that is typically applied. Furthermore, the high particle flux environment, in combination with the self-triggered readout system, required that only data from hit channels be recorded, as opposed to the full set of readout channels. This setup differs significantly from the expected conditions in BabyIAXO, where typical event rates are below 1\,Hz. These constraints introduced additional limitations in the evaluation of gas diffusion and spatial resolution, which are reflected in the reported measurement uncertainties.

The detector was tested at different drift field values (from 50\,V/cm to 400\,V/cm) and under different beam energies (from 5\,keV to 10\,keV). The beam sizes were adjusted to several values between 90 \,$\times$ 90\,$\microns^2$ to 1.3 $\times$ 1.3\,mm$^2$. The smallest beam size was chosen for the spatial resolution study in the center of the detector. For the position scan of the spatial resolution, different beam sizes were tested. A summary of the measurements performed is given in Table~\ref{tab:measurements}.

\begin{table}[ht]
    \centering
    \begin{tabular}{|c|c|}
        \hline
        & Energy scan \\
        \hline
        Position &  Centre \\
        Beam size & 90\,$\times$90\,$\microns^2$ to 1.3\,$\times$1.3\,mm$^2$ \\
        Energy & 5\,keV to 10\,keV \\
        Drift field & 100\,V/cm\\
        \hline
        \hline
        & Drift field scan \\
        \hline
        Position &  Centre \\
        Beam size & 90\,$\times$90\,$\microns^2$ \\
        Energy & 5\,keV to 10\,keV \\
        Drift field & 50\,V/cm to 400\,V/cm\\
        \hline
        \hline
        & Position scan \\
        \hline
        Position X &  -7\, to 9\,mm \\
        Position Y &  -7\, to 7\,mm  \\
        Beam size & 90\,$\times$90\,$\microns^2$ to 1.3\,$\times$1.3\,mm$^2$ \\
        Energy & 6\,keV \\
        Drift field & 100\,V/cm\\
        \hline
    \end{tabular}
    \caption{Summary table of the measurements taken during the test. Four beam sizes were measured 90 $\times$ 90, 200 $\times$ 200, 900 $\times$ 900 and 1300$\times$ 1300 $\microns^2$. The X-ray energies were varied across five values: 5, 6, 7, 8.5, and 10\,keV.}
    \label{tab:measurements}
\end{table}

%%%%%%%%%%%%
\section{Data processing and analysis}
\label{section:analysis}
%%%%%%%%%%%%
The data acquired with the Microbulk detector are processed and analyzed using the REST-for-Physics software framework~\citep{REST}. This software processes the signals from each strip extracting the spectra and the 3-dimensional information of the charge deposition.

%\textbf{This is a copy from another paper:} \textit{The data is processed in three steps using the appropriate REST-for-Physics libraries: during the raw signal analysis noise events are identified and removed, and the signal pulses are isolated; in the detector hits analysis the signals from electronic channels are translated to energy deposits (``hits") and associated with physical coordinates (X,Y) in the readout geometry; the size of the energy deposit in the Z-coordinate is calculated from the charge collection time and drift velocity, which in turn is calculated with the Garfield++ code~\citep{garfield}; finally, in the track analysis the energy deposits are connected to 3-dimensional tracks by using multiple algorithms to find the shortest path that interconnects energy deposits within an event, adding another level of topological information. The processing generates many observables -- per-event information extracted from the data -- that are used to define selection criteria for background discrimination.}

The data are processed in three sequential steps using the REST-for-Physics framework. In the initial stage, raw signal analysis, noise events are identified and removed, and individual signal pulses are isolated. In the subsequent detector hits analysis, signals from the electronic channels are translated into energy deposits (“hits”) and assigned physical coordinates (X, Y) within the readout geometry. The extent of the energy deposit along the Z-axis is inferred from the charge collection time, using a drift velocity calculated with the Garfield++ simulation package~\citep{garfield}. In the final stage, track reconstruction, the identified energy deposits are connected into three-dimensional tracks. This is achieved through a series of algorithms designed to determine the shortest paths interlinking hits within an event, thereby providing additional topological information. This data processing workflow yields a broad set of observables—per-event quantities extracted from the recorded signals—which are subsequently used to define selection criteria for background rejection and signal discrimination.

\begin{figure}[t]
    \centering
    \begin{minipage}[b]{0.5\textwidth}
    \includegraphics[height=5.1cm]{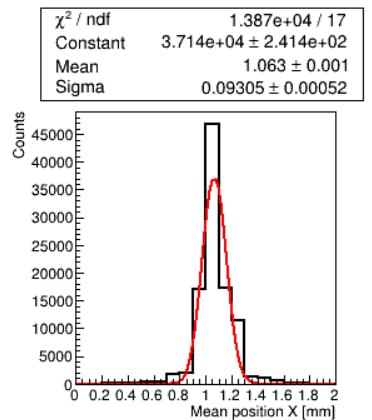}
    \end{minipage}
    \hfill
    \begin{minipage}[b]{0.45\textwidth}
    \includegraphics[height=5.12cm]{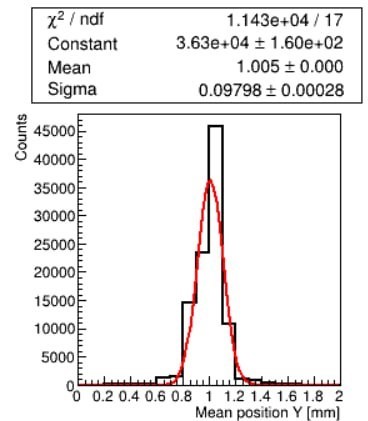}
    \end{minipage}
    \caption{Histograms of the observable of the mean position in X(Y) and its gaussian fit for the case of 6\,keV, a beam size of 90\,$\times$90\,$\microns^2$ and a 100\,V/cm drift in the detector chamber.}
    \label{fig:mean_fit_6keV}
    \begin{minipage}[b]{0.5\textwidth}
    \includegraphics[height=5.1cm]{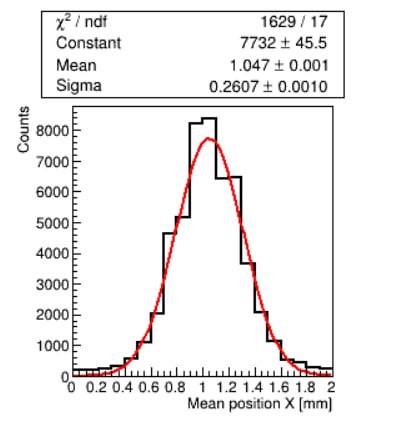}
    \end{minipage}
    \hfill
    \begin{minipage}[b]{0.45\textwidth}
    \includegraphics[height=5.1cm]{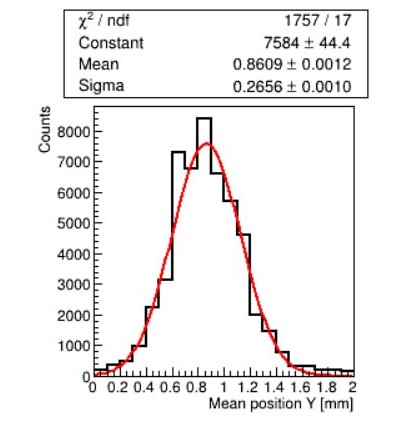}
    \end{minipage}
    \caption{Histograms of the observable of the mean position in X(Y) and its gaussian fit for the case of 10\,keV, a beam size of 90\,$\times$90\,$\microns^2$ and a 100\,V/cm drift in the detector chamber.}
    \label{fig:mean_fit_10keV}
\end{figure}

For the purpose of this analysis, the only cuts applied for the event selection were:
\begin{enumerate}
    \item[-] Minimum hit requirement: Events were required to have at least one hit in both the X and Y coordinates. This standard pre-selection step is used so that the event can be assigned a readout position and to eliminate possible non-physical or spurious events.
    \item [-] Single-track selection: Only events reconstructing a single track were retained. X-rays photoabsorption is a point-like event that results in a single-track signature. %in contrast to background events like charged particle interactions that usually leave multiple tracks. This criterion effectively isolates X-ray events, 
    By applying this cut, approximately \SI{96}{\%} of the total calibration events are selected.
\end{enumerate}

The spatial resolution is estimated from the mean position of the events. Since the size of the beam chosen is sufficiently small and the X-rays are point-like events in the IAXO-D1 detector, the resolution can be estimated as the standard deviation ($\sigma$) of the mean position observable from a gaussian fit of the histogram as shown in figures \ref{fig:mean_fit_6keV} and \ref{fig:mean_fit_10keV}. Alternative fit models were also explored, including a step function convoluted with a Gaussian—to account for the square shape of the beam. However, no significant differences were observed in the extracted resolution, suggesting that the detector resolution is of the order of the beam size. %The bin size can also affect the estimation of the resolution, by checking the behavior for larger beams it has been observed that the limiting factor is not the bin size but the intrinsic resolution of the detector (and the beam).

%%%%%%%%%%%%
\section{Simulations}
\label{section:Simulation}
%%%%%%%%%%%%
The study has been complemented with extensive radiation transport simulations using Geant4~\citep{Geant4_2016}. REST-for-Physics has been used as the software framework to interface with Geant4, and for processing the resulting data~\citep{REST}. The model used includes the detector chamber and the Micromegas readout. % (see figure \ref{fig:simu}). 
The libraries from REST-for-physics allow to change from the Geant4 energy deposits to the ones that would be produced by the physical setup mimicking the raw signals. To achieve a faithful simulation of the data, relevant physical processes such as electron drift and diffusion within the gas, along with realistic noise conditions and energy thresholds observed during data acquisition, are incorporated. These simulated signals are then subjected to the same reconstruction and analysis pipeline as the experimental data, enabling a direct and meaningful comparison between the two.

For the simulation setup, events are generated just above the Mylar entrance window, with an initial direction perpendicular to the readout plane. The spatial distribution of the beam is modeled as a square profile, consistent with the size and shape of the beam measured by the reference camera.

%\begin{figure}[h]
%	\centering
%    \includegraphics[height=7.5 cm]{Screenshot from 2024-10-09 10-59-19.png}
%	\caption{Simulation of a gamma event generated at the entrance window of the detector in a perpendicular direction (as coming from the beam).}
%    \label{fig:simu}
%\end{figure}

For the purpose of this study, the simulations were carried out under two distinct operating conditions: the high noise and elevated threshold environment characteristics of the SOLEIL beamline, and the more favorable noise and threshold conditions typically encountered during laboratory operation of the detector. For these two scenarios, the energy thresholds were set to the values mentioned in section \ref{section:Setup}, and the noise was added as a gaussian oscillation from the ideal signal with the sigma measured from the baseline during the test and in the laboratory.

%%%%%%%%%%%%%%%%%%%%%%%%%%%%%%%%%%%%%%%%%%%%%%%%%%%%%%
\section{Results}
\label{sec:results}
%%%%%%%%%%%%%%%%%%%%%%%%%%%%%%%%%%%%%%%%%%%%%%%%%%%%%%

This section presents the main results of the spatial resolution study conducted at the SOLEIL synchrotron using the IAXO-D1 detector setup described in Section~\ref{section:Setup}. The resolution is estimated from the mean position observable, as discussed in section \ref{section:analysis}. Simulated data were produced as described in section \ref{section:Simulation}, for each data point a sample of $\text{10}^\text{5}$ events were generated.

\subsection{Resolution as a function of energy.}
\begin{figure}[t]
  \centering
  \begin{minipage}[b]{0.45\textwidth}
    \includegraphics[width=5.8cm]{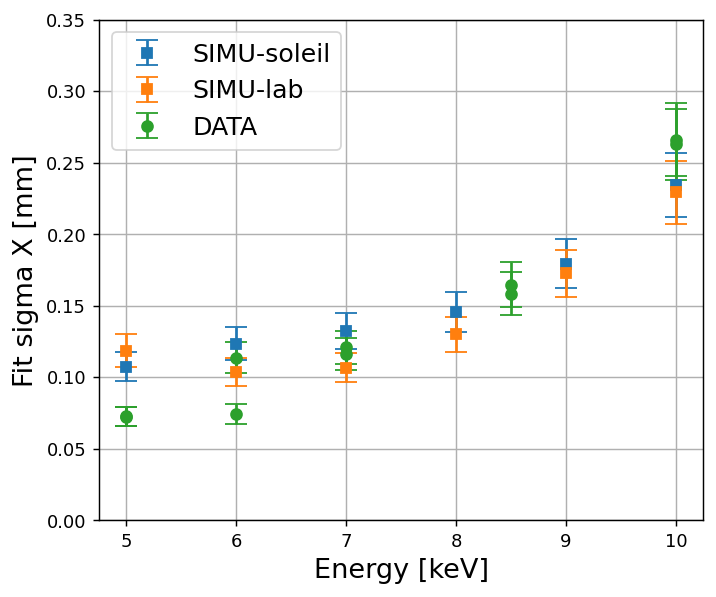}
  \end{minipage}
  \hfill
  \begin{minipage}[b]{0.5\textwidth}
    \includegraphics[width=5.8cm]{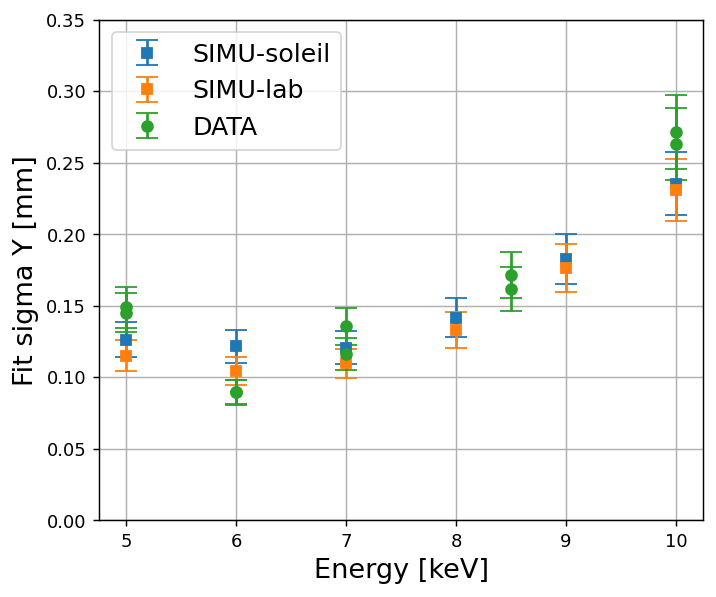}
  \end{minipage}
  \caption{Resolution X(Y) vs energy at 100V/cm. The error bars are $\times$30 the statistical error of the sigma estimation from the gaussian fit.}
  \label{fig:res-E}
\end{figure}

To assess the spatial resolution of the IAXO-D1 detector, its response to different beam energies was measured, as shown in figure~\ref{fig:res-E}. The detector was operated at nominal voltage settings: mesh voltage of \SI{360}{V} and a drift field of \SI{100}{V/cm}. The X-ray beam size was fixed at 90\,$\times$90\,$\microns^2$, the smallest one explored, to ensure that the estimate of the resolution, as the standard deviation of the mean position, serves as a good approximation of the detector's resolution.

Figure \ref{fig:res-E} shows that the experimental results are in good agreement with the simulated data under both SOLEIL and laboratory conditions. The differences between these two simulated scenarios are visible, supporting the hypothesis that laboratory conditions would result in an improvement of the measurement of the spatial resolution. The resolution exhibits symmetric behavior in the X and Y directions, with a minimum slightly under $\text{100}$\,$\microns$ at 6\,keV and higher values for increasing energies reaching $\sim\text{260}$\,$\microns$ at 10\,keV. This degradation observed at higher energies is associated with longer photoelectron tracks produced in the gas mixture, leading to increased charge spread, reducing the precision in position reconstruction \cite{Cools:2024cjl}. The observation of the track size influence on the sigma indicates that the intrinsic resolution of the detector is at least as good as the lowest value measured.

\begin{figure}[h]
  \centering
  \begin{minipage}[b]{0.45\textwidth}
    \includegraphics[width=5.8cm]{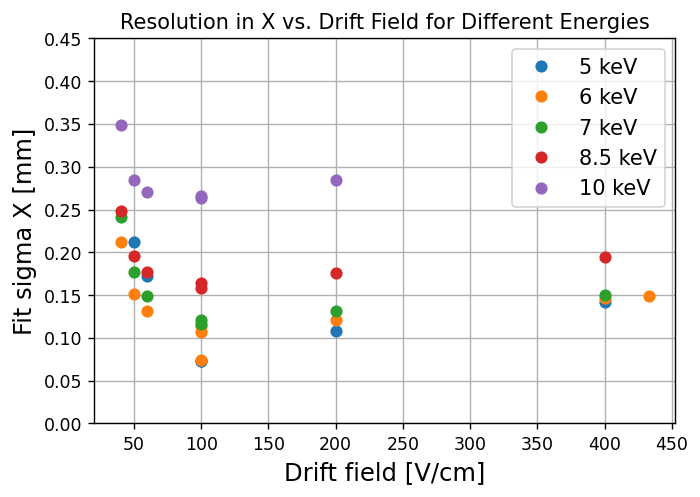}
  \end{minipage}
  \hfill
  \begin{minipage}[b]{0.5\textwidth}
    \includegraphics[width=5.8cm]{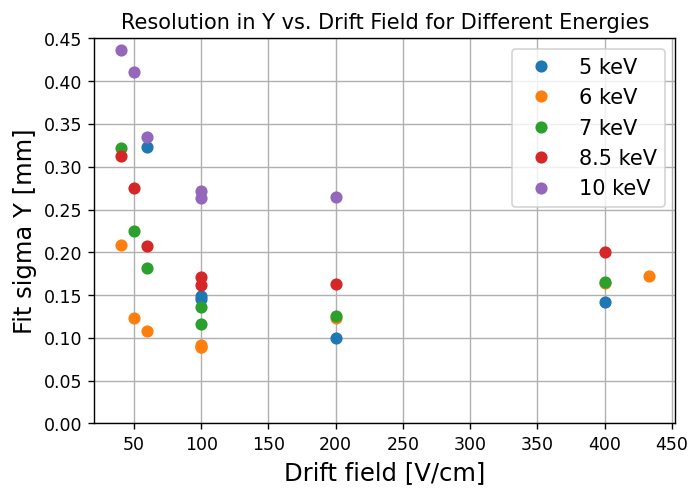}
  \end{minipage}
  \caption{Resolution X(Y) vs drift field, for different energies. The error bars are not represented on the plots in order to facilitate the visualization of all the values shown. The uncertainty of the fit sigmas is the order of 2-5\,\%.}
  \label{fig:res-Ed}
\end{figure}
\subsection{Resolution as a function of drift field.}
The IAXO-D1 detector response was also studied for several values of the electric field in its drift region by setting the voltage of the cathode. The spatial resolution as a function of the drift field was measured for various X-ray energies and is summarised in figure \ref{fig:res-Ed}. The behavior is consistent for all the energies tested: the resolution reaches a minimum value around 100\,V/cm, degrades rapidly at lower drift fields, and increases gradually for higher intensities of the drift. For all the cases shown, the spatial resolution at lower energies is better.

\begin{figure}[h!]
  \centering
  \begin{minipage}[b]{0.45\textwidth}
    \includegraphics[width=5.8cm]{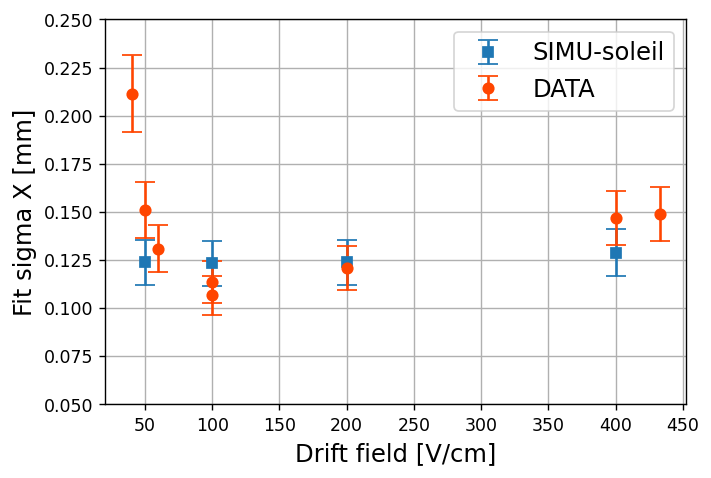}
  \end{minipage}
  \hfill
  \begin{minipage}[b]{0.5\textwidth}
    \includegraphics[width=5.8cm]{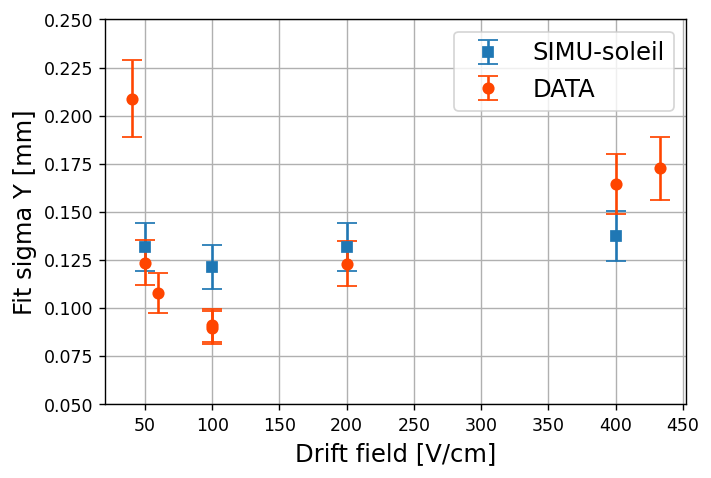}
  \end{minipage}
  \begin{minipage}[b]{0.45\textwidth}
    \includegraphics[width=5.8cm]{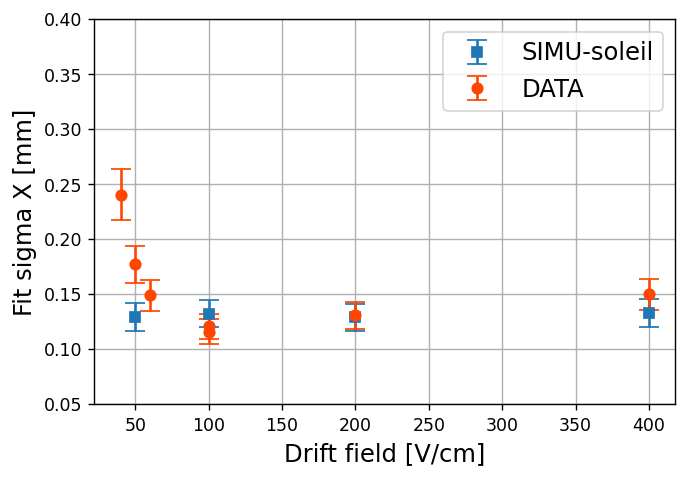}
  \end{minipage}
  \hfill
  \begin{minipage}[b]{0.5\textwidth}
    \includegraphics[width=5.8cm]{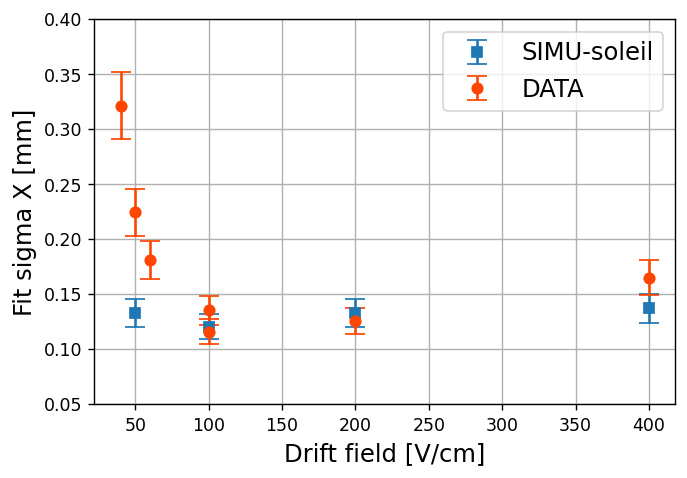}
  \end{minipage}
  \caption{Resolution X(Y) vs drift field at 6\,keV (upper plots) and 7\,keV (lower plots), and its comparison with the simulated data. The error bars are $\times$30 the statistical error of the sigma estimation from the gaussian fit.}
  \label{fig:res-Ed_E6-7}
\end{figure}

The dependence of spatial resolution on the drift field was also compared with simulated data. As representative examples, the cases of 6\,keV and 7\,keV are shown in figure \ref{fig:res-Ed_E6-7}. The experimental data follow the expected behavior for drifts above 100\,V/cm, although, for lower drift fields, there is a noticeable discrepancy.
This mismatch can be explained by the presence of impurities in the gas, such as O$_2$ or H$_2$O, which became more relevant at low drift field, reducing the electron collection efficiency, an effect not implemented in our simulation. The gradual deterioration of the resolution at higher drifts is attributed to the increase in transverse diffusion.

\subsection{Resolution as a function of beam detector position.}
\begin{figure}[ht]
    \centering
    \includegraphics[width=0.35\linewidth]{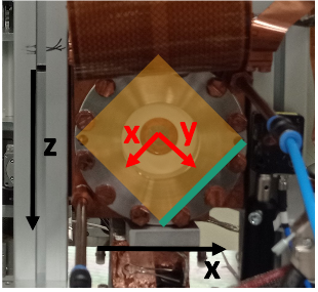}
    \caption{Front view of the detector in the beam line. Red: the coordinates of the readout. Black: coordinates of the platform. Green: position of the HV conections inside the detector chamber.}
    \label{fig:positions}
\end{figure}

To conclude with the study of the spatial resolution, various positions of the IAXO-D1 readout plane were measured by moving the detector platform along its X and Z axis. The results presented correspond to a beam of 90$\times$ 90 $\microns^2$ and 6\,keV, and a drift field of 100\,V/cm. The displacement is illustrated in figure \ref{fig:positions} where the coordinates of the platform and readout are depicted. In order to facilitate the interpretation of the spatial resolution, the results will be presented as a function of the mean position in that same axis. Before proceeding to the discussion, it is important to recall that the field shapers were not operational during the tests conducted at SOLEIL.
\begin{figure}[ht]
  \centering
  \begin{minipage}[b]{0.45\textwidth}
    \includegraphics[width=5.8cm]{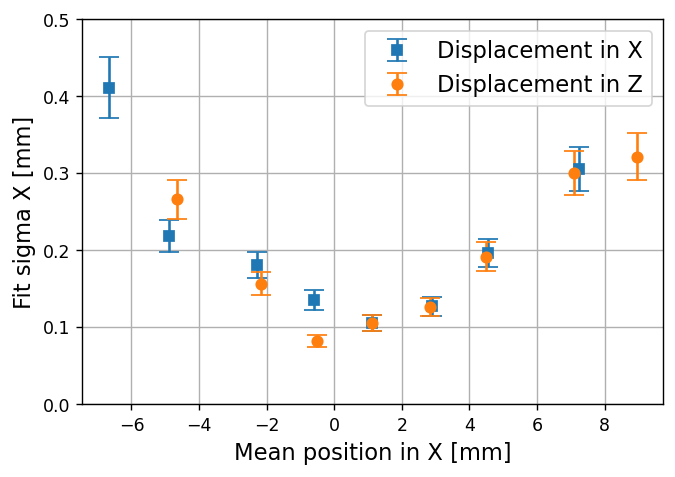}
  \end{minipage}
  \hfill
  \begin{minipage}[b]{0.5\textwidth}
    \includegraphics[width=5.8cm]{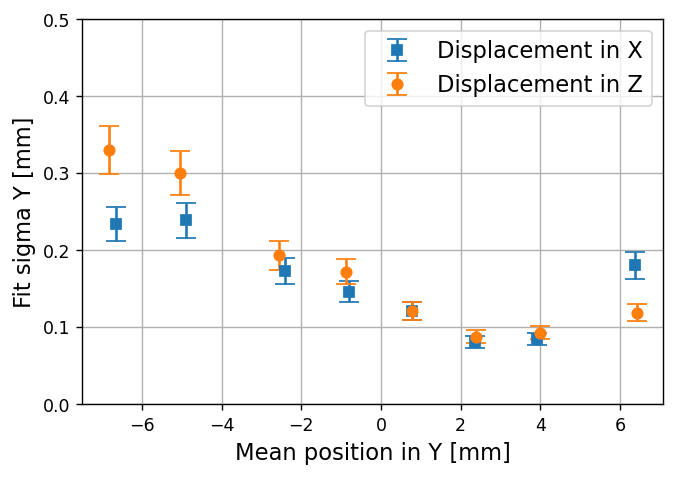}
  \end{minipage}
  \caption{ResolutionX(Y) vs the position along the axis. The error bars are $\times$30 the statistical error of the sigma estimation from the gaussian fit. Blue: values obtained when the detector was being moved in X. Orange: values obtained when the detector was being moved in Z.}
  \label{fig:res-pos}
\end{figure}

The measured spatial resolution as a function of the position is shown in figure \ref{fig:res-pos}. Along the X axis, the resolution exhibits a minimum of about 100\,\microns at the center of the detector, as expected due to the symmetry of the electric field. However, this is not the case for the Y axis, where the same minimum is observed at approximately Y$ \sim 2\,\mathrm{mm}$. This asymmetry is attributed to the drift field caused by the high-voltage connections located on one side of the readout, perpendicular to the Y axis (as illustrated in figure \ref{fig:positions}). These field distortions were not investigated further in the scope of this study.

In figure \ref{fig:res-pos} is shown that, when placing the beam just a few millimeters away from the center, a noticeable degradation in resolution by a factor 3-4 is observed. This emphasizes the importance of the field shaper rings, even within the fiducial radius smaller than $1\,\mathrm{cm}$ from the center. Although the effect in the position is significant, the spatial resolution is well below the needed constrain of \SI{1}{mm}.

%%%%%%%%%%%%%%%%%%%%%%%%%%%%%%%%%%%%%%%%%%%%%%%%%%%%%%
\section{Conclusions and perspectives}
\label{section:Conclusion}
%%%%%%%%%%%%%%%%%%%%%%%%%%%%%%%%%%%%%%%%%%%%%%%%%%%%%%
The use of SOLEIL’s X-ray beam allowed a detailed characterization of the IAXO-D1 two-dimensional Micromegas detector. At the detector center, the spatial resolution measured was 100-300\,\microns, in the range between 5\,keV and 10\,keV, when operating at a drift field of at least \SI{100}{V/cm}. The results are in good agreement with the dedicated simulations performed. These values represent conservative estimates, as discussed throughout the article, the beam size is comparable to the measured resolution. Therefore, the extracted values include contributions from both the intrinsic detector resolution and the finite beam width. Inhomogeneities in the fiducial region of \SI{1}{cm} radius from the center have been observed to cause a degradation of the spatial resolution, while its worst value was still well below the \SI{1}{mm} constraint on BabyIAXO experiment. Small effect on the spatial resolution due to higher noise conditions were also verified with simulation. This study experimentally demonstrates that Micromegas technology fulfills and exceeds the required BabyIAXO performance specifications.

The measured spatial resolution positions the 2D-Microbulk technology as a strong candidate for neutron imaging~\cite{jimaging3040064,book, Pancin:2004dp, Jeanneau:2004mlk}, where high spatial precision and minimal material thickness are critical requirements.

%Possible future test with the field shaper operative and after working on the noise conditions.

%Background measurements in Canfranc, Saclay, Zaragoza with different shieldings and veto configurations.
%Soon to come installation in DESY for an in situ background data taking.

%%%%%%%%%%%%%%%%%%%%%%%
\section*{Acknowledgments}
%%%%%%%%%%%%%%%%%%%%%%%
The authors acknowledge support from the Agence Nationale de la Recherche (France) ANR-19-CE31-0024, from the European Research Council (ERC) under the European Union’s Horizon 2020 research and innovation programme (ERC-2017-AdG IAXO+, grant agreement No. 788781),   from the Agencia Estatal de Investigación (AEI) under the grant agreement PID2022-137268NB-C51 funded by MCIN/AEI/10.13039/501100011033/FEDER, as well as funds from “European Union NextGenerationEU/PRTR” (Planes complementarios, Programa de Astrofísica y Física de Altas Energías).
The authors acknowledge SOLEIL for provision of synchrotron radiation facilities (proposal number 20221337) and we would like to thank Pascal Mercere and Paulo Da Silva for assistance in using METROLOGIE beamline.
\newpage
\bibliography{mybibfile}

\begin{thebibliography}{10}
\expandafter\ifx\csname url\endcsname\relax
  \def\url#1{\texttt{#1}}\fi
\expandafter\ifx\csname urlprefix\endcsname\relax\def\urlprefix{URL }\fi
\expandafter\ifx\csname href\endcsname\relax
  \def\href#1#2{#2} \def\path#1{#1}\fi

\bibitem{IAXO:2020wwp}
A.~Abeln, et~al., {Conceptual design of BabyIAXO, the intermediate stage towards the International Axion Observatory}, JHEP 05 (2021) 137.
\newblock \href {http://arxiv.org/abs/2010.12076} {\path{arXiv:2010.12076}}, \href {https://doi.org/10.1007/JHEP05(2021)137} {\path{doi:10.1007/JHEP05(2021)137}}.

\bibitem{IAXO:2024wss}
S.~Ahyoune, et~al., {An accurate solar axions ray-tracing response of BabyIAXO}, JHEP 02 (2025) 159.
\newblock \href {http://arxiv.org/abs/2411.13915} {\path{arXiv:2411.13915}}, \href {https://doi.org/10.1007/JHEP02(2025)159} {\path{doi:10.1007/JHEP02(2025)159}}.

\bibitem{IAXO:2019mpb}
E.~Armengaud, et~al., {Physics potential of the International Axion Observatory (IAXO)}, JCAP 06 (2019) 047.
\newblock \href {http://arxiv.org/abs/1904.09155} {\path{arXiv:1904.09155}}, \href {https://doi.org/10.1088/1475-7516/2019/06/047} {\path{doi:10.1088/1475-7516/2019/06/047}}.

\bibitem{Carenza:2024ehj}
P.~Carenza, M.~Giannotti, J.~Isern, A.~Mirizzi, O.~Straniero, {Axion astrophysics}, Phys. Rept. 1117 (2025) 1--102.
\newblock \href {http://arxiv.org/abs/2411.02492} {\path{arXiv:2411.02492}}, \href {https://doi.org/10.1016/j.physrep.2025.02.002} {\path{doi:10.1016/j.physrep.2025.02.002}}.

\bibitem{Giomataris:1995fq}
Y.~Giomataris, P.~Rebourgeard, J.~P. Robert, G.~Charpak, {MICROMEGAS: A High granularity position sensitive gaseous detector for high particle flux environments}, Nucl. Instrum. Meth. A376 (1996) 29--35.
\newblock \href {https://doi.org/10.1016/0168-9002(96)00175-1} {\path{doi:10.1016/0168-9002(96)00175-1}}.

\bibitem{Andriamonje:2010zz}
S.~Andriamonje, et~al., {Development and performance of Microbulk Micromegas detectors}, JINST 5 (2010) P02001.
\newblock \href {https://doi.org/10.1088/1748-0221/5/02/P02001} {\path{doi:10.1088/1748-0221/5/02/P02001}}.

\bibitem{CAST:2004gzq}
K.~Zioutas, et~al., {First results from the CERN Axion Solar Telescope (CAST)}, Phys. Rev. Lett. 94 (2005) 121301.
\newblock \href {http://arxiv.org/abs/hep-ex/0411033} {\path{arXiv:hep-ex/0411033}}, \href {https://doi.org/10.1103/PhysRevLett.94.121301} {\path{doi:10.1103/PhysRevLett.94.121301}}.

\bibitem{CAST:2008ixs}
E.~Arik, et~al., {Probing eV-scale axions with CAST}, JCAP 02 (2009) 008.
\newblock \href {http://arxiv.org/abs/0810.4482} {\path{arXiv:0810.4482}}, \href {https://doi.org/10.1088/1475-7516/2009/02/008} {\path{doi:10.1088/1475-7516/2009/02/008}}.

\bibitem{CAST:2011rjr}
S.~Aune, et~al., {CAST search for sub-eV mass solar axions with 3He buffer gas}, Phys. Rev. Lett. 107 (2011) 261302.
\newblock \href {http://arxiv.org/abs/1106.3919} {\path{arXiv:1106.3919}}, \href {https://doi.org/10.1103/PhysRevLett.107.261302} {\path{doi:10.1103/PhysRevLett.107.261302}}.

\bibitem{CAST:2017uph}
V.~Anastassopoulos, et~al., {New CAST Limit on the Axion-Photon Interaction}, Nature Phys. 13 (2017) 584--590.
\newblock \href {http://arxiv.org/abs/1705.02290} {\path{arXiv:1705.02290}}, \href {https://doi.org/10.1038/nphys4109} {\path{doi:10.1038/nphys4109}}.

\bibitem{CAST:2024eil}
K.~Altenm\"uller, et~al., {New Upper Limit on the Axion-Photon Coupling with an Extended CAST Run with a Xe-Based Micromegas Detector}, Phys. Rev. Lett. 133~(22) (2024) 221005.
\newblock \href {http://arxiv.org/abs/2406.16840} {\path{arXiv:2406.16840}}, \href {https://doi.org/10.1103/PhysRevLett.133.221005} {\path{doi:10.1103/PhysRevLett.133.221005}}.

\bibitem{Altenmuller:2024uza}
K.~Altenm{\"u}ller, et~al., {Background discrimination with a Micromegas detector prototype and veto system for BabyIAXO}, Front. in Phys. 12 (2024) 1384415.
\newblock \href {http://arxiv.org/abs/2403.06316} {\path{arXiv:2403.06316}}, \href {https://doi.org/10.3389/fphy.2024.1384415} {\path{doi:10.3389/fphy.2024.1384415}}.

\bibitem{Hoof:2023jol}
S.~Hoof, J.~Jaeckel, L.~J. Thormaehlen, {Axion helioscopes as solar thermometers}, JCAP 10 (2023) 024.
\newblock \href {http://arxiv.org/abs/2306.00077} {\path{arXiv:2306.00077}}, \href {https://doi.org/10.1088/1475-7516/2023/10/024} {\path{doi:10.1088/1475-7516/2023/10/024}}.

\bibitem{Aznar:2015iia}
F.~Aznar, et~al., {A Micromegas-based low-background x-ray detector coupled to a slumped-glass telescope for axion research}, JCAP 12 (2015) 008.
\newblock \href {http://arxiv.org/abs/1509.06190} {\path{arXiv:1509.06190}}, \href {https://doi.org/10.1088/1475-7516/2015/12/008} {\path{doi:10.1088/1475-7516/2015/12/008}}.

\bibitem{MirallasSanchez:2024bcp}
H.~Mirallas~S{\'a}nchez, {Desarrollo de grandes planos de lectura Micromegas para experimentos de b{\'u}squeda de sucesos poco probables}, Ph.D. thesis, U. Zaragoza (main) (2024).

\bibitem{Baron:2017kld}
P.~Baron, D.~Calvet, F.~Château, A.~Corsi, E.~Delagnes, A.~Delbart, A.~Obertelli, N.~Paul, {Operational Experience With the Readout System of the MINOS Vertex Tracker}, IEEE Trans. Nucl. Sci. 64~(6) (2017) 1494--1500.
\newblock \href {https://doi.org/10.1109/TNS.2017.2706971} {\path{doi:10.1109/TNS.2017.2706971}}.

\bibitem{ElisaThesis}
E.~Ruiz~Ch\'{o}liz, {Ultra-low background Micromegas X-ray detectors for Axion searches in IAXO and BabyIAXO}, {PhD thesis}, Universidad de Zaragoza (2019).

\bibitem{Metrology1}
M.~Idir, P.~Mercere, T.~Moreno, A.~Delmotte, \href{https://aip.scitation.org/doi/abs/10.1063/1.2436137}{{Metrology and Tests Beamline at SOLEIL}}, AIP Conference Proceedings 879~(1) (2007) 619--622.
\newblock \href {http://arxiv.org/abs/https://aip.scitation.org/doi/pdf/10.1063/1.2436137} {\path{arXiv:https://aip.scitation.org/doi/pdf/10.1063/1.2436137}}, \href {https://doi.org/10.1063/1.2436137} {\path{doi:10.1063/1.2436137}}.
\newline\urlprefix\url{https://aip.scitation.org/doi/abs/10.1063/1.2436137}

\bibitem{Metrology2}
Y.~Ménesguen, M.-C. Lépy, \href{https://analyticalsciencejournals.onlinelibrary.wiley.com/doi/abs/10.1002/xrs.1366}{{Characterization of the Metrology beamline at the SOLEIL synchrotron and application to the determination of mass attenuation coefficients of Ag and Sn in the range 3.5$\leq$E$\leq$28 keV}}, X-Ray Spectrometry 40~(6) (2011) 411--416.
\newblock \href {https://doi.org/https://doi.org/10.1002/xrs.1366} {\path{doi:https://doi.org/10.1002/xrs.1366}}.
\newline\urlprefix\url{https://analyticalsciencejournals.onlinelibrary.wiley.com/doi/abs/10.1002/xrs.1366}

\bibitem{baslerCam}
\href{{http://www.altavision.com.br/Datasheets/Basler_EN/ scA1300-32gm.htmlx}}{Altavision}.
\newline\urlprefix\url{{http://www.altavision.com.br/Datasheets/Basler_EN/ scA1300-32gm.htmlx}}

\bibitem{REST}
K.~Altenmüller, S.~Cebrián, T.~Dafni, D.~Díez-Ibáñez, J.~Galán, J.~Galindo, J.~A. García, I.~G. Irastorza, G.~Luzón, C.~Margalejo, H.~Mirallas, L.~Obis, O.~Pérez, K.~Han, K.~Ni, Y.~Bedfer, B.~Biasuzzi, E.~Ferrer-Ribas, D.~Neyret, T.~Papaevangelou, C.~Cogollos, E.~Picatoste, \href{https://www.sciencedirect.com/science/article/pii/S0010465521003933}{{REST-for-Physics}, a {ROOT}-based framework for event oriented data analysis and combined monte carlo response}, Computer Physics Communications 273 (2022) 108281.
\newblock \href {https://doi.org/https://doi.org/10.1016/j.cpc.2021.108281} {\path{doi:https://doi.org/10.1016/j.cpc.2021.108281}}.
\newline\urlprefix\url{https://www.sciencedirect.com/science/article/pii/S0010465521003933}

\bibitem{garfield}
H.~Schindler, R.~Veenhof, {Garfield++ — simulation of ionisation based tracking detectors}\url{http://garfieldpp.web.cern.ch/garfieldpp}, accessed: 2024-04-02 (2024).

\bibitem{Geant4_2016}
J.~Allison, K.~Amako, J.~Apostolakis, P.~Arce, M.~Asai, T.~Aso, E.~Bagli, A.~Bagulya, S.~Banerjee, G.~Barrand, B.~Beck, A.~Bogdanov, D.~Brandt, J.~Brown, H.~Burkhardt, P.~Canal, D.~Cano-Ott, S.~Chauvie, K.~Cho, G.~Cirrone, G.~Cooperman, M.~Cortés-Giraldo, G.~Cosmo, G.~Cuttone, G.~Depaola, L.~Desorgher, X.~Dong, A.~Dotti, V.~Elvira, G.~Folger, Z.~Francis, A.~Galoyan, L.~Garnier, M.~Gayer, K.~Genser, V.~Grichine, S.~Guatelli, P.~Guèye, P.~Gumplinger, A.~Howard, I.~Hřivnáčová, S.~Hwang, S.~Incerti, A.~Ivanchenko, V.~Ivanchenko, F.~Jones, S.~Jun, P.~Kaitaniemi, N.~Karakatsanis, M.~Karamitros, M.~Kelsey, A.~Kimura, T.~Koi, H.~Kurashige, A.~Lechner, S.~Lee, F.~Longo, M.~Maire, D.~Mancusi, A.~Mantero, E.~Mendoza, B.~Morgan, K.~Murakami, T.~Nikitina, L.~Pandola, P.~Paprocki, J.~Perl, I.~Petrović, M.~Pia, W.~Pokorski, J.~Quesada, M.~Raine, M.~Reis, A.~Ribon, A.~R. Fira, F.~Romano, G.~Russo, G.~Santin, T.~Sasaki, D.~Sawkey, J.~Shin, I.~Strakovsky, A.~Taborda, S.~Tanaka, B.~Tomé, T.~Toshito, H.~Tran,
  P.~Truscott, L.~Urban, V.~Uzhinsky, J.~Verbeke, M.~Verderi, B.~Wendt, H.~Wenzel, D.~Wright, D.~Wright, T.~Yamashita, J.~Yarba, H.~Yoshida, \href{https://www.sciencedirect.com/science/article/pii/S0168900216306957}{{Recent developments in Geant4}}, Nuclear Instruments and Methods in Physics Research Section A: Accelerators, Spectrometers, Detectors and Associated Equipment 835 (2016) 186--225.
\newblock \href {https://doi.org/https://doi.org/10.1016/j.nima.2016.06.125} {\path{doi:https://doi.org/10.1016/j.nima.2016.06.125}}.
\newline\urlprefix\url{https://www.sciencedirect.com/science/article/pii/S0168900216306957}

\bibitem{Cools:2024cjl}
A.~Cools, E.~Ferrer-Ribas, T.~Papaevangelou, E.~C. Pollacco, M.~Lisowska, F.~M. Brunbauer, E.~Oliveri, F.~J. Iguaz, {Spatial resolution studies using point spread function extraction in optically read out Micromegas and GEM detectors}, Nucl. Instrum. Meth. A 1069 (2024) 169933.
\newblock \href {http://arxiv.org/abs/2407.15491} {\path{arXiv:2407.15491}}, \href {https://doi.org/10.1016/j.nima.2024.169933} {\path{doi:10.1016/j.nima.2024.169933}}.

\bibitem{jimaging3040064}
M.~Strobl, R.~P. Harti, C.~Gruenzweig, R.~Woracek, J.~Plomp, \href{https://www.mdpi.com/2313-433X/3/4/64}{Small angle scattering in neutron imaging—a review}, Journal of Imaging 3~(4) (2017).
\newblock \href {https://doi.org/10.3390/jimaging3040064} {\path{doi:10.3390/jimaging3040064}}.
\newline\urlprefix\url{https://www.mdpi.com/2313-433X/3/4/64}

\bibitem{book}
I.~Anderson, R.~McGreevy, H.~Bilheux, Neutron Imaging and Applications, 2009.
\newblock \href {https://doi.org/10.1007/978-0-387-78693-3} {\path{doi:10.1007/978-0-387-78693-3}}.

\bibitem{Pancin:2004dp}
J.~Pancin, et~al., {Measurement of the n{\_}TOF beam profile with a micromegas detector}, Nucl. Instrum. Meth. A 524 (2004) 102--114.
\newblock \href {https://doi.org/10.1016/j.nima.2004.01.055} {\path{doi:10.1016/j.nima.2004.01.055}}.

\bibitem{Jeanneau:2004mlk}
F.~Jeanneau, et~al., {Neutron imaging with a Micromegas detector}, IEEE Trans. Nucl. Sci. 53 (2006) 595--600.
\newblock \href {http://arxiv.org/abs/physics/0607191} {\path{arXiv:physics/0607191}}, \href {https://doi.org/10.1109/TNS.2006.870175} {\path{doi:10.1109/TNS.2006.870175}}.

\end{thebibliography}
\bibliographystyle{elsarticle-num}
\end{document}